\documentclass[twocolumn,preprintnumbers,floatfix,superscriptaddress,nofootinbib]{revtex4}
\usepackage[utf8]{inputenc}
\usepackage{setspace}
\usepackage{booktabs}
\usepackage{rotating}
\usepackage{url,comment}
\usepackage{times}
\usepackage{latexsym}
\usepackage{graphicx, graphics, amssymb, slashed, xcolor,amsthm, array}
\usepackage{subfigure}
\usepackage{listings} 
\usepackage{afterpage}
\usepackage{physics}
\usepackage[colorlinks=true,linkcolor=blue,citecolor=blue,urlcolor=blue]{hyperref}
\usepackage{here}
\usepackage{MnSymbol} 
\usepackage{braket}

\newcommand\eea{\end{aligned} \end{equation}}
\newcommand\bea{\begin{equation} \begin{aligned}}
\newcommand\UI{U_{\text{infer}}}
\newcommand\UGR{U_{\text{GR}}}

\begin{document}

%\vspace*{-30mm}

\title{Searching for a Fifth Force with Atomic and Nuclear Clocks}

%%%%%%%%%%%%%%%%%%%%%%%%%%%%%%%%%%%%%%%%%%%%%%%%%%%%%%%

\author{Dawid Brzeminski}
\email{dbrzemin@umd.edu}
\affiliation{Maryland Center for Fundamental Physics, Department of Physics, University of Maryland, College Park, MD 20742, U.S.A.}

\author{Zackaria Chacko}
\email{zchacko@umd.edu}
\affiliation{Maryland Center for Fundamental Physics, Department of Physics, University of Maryland, College Park, MD 20742, U.S.A.}

\author{Abhish Dev}
\email{abhish@fnal.gov}
\affiliation{Theoretical Physics Department, Fermilab, P.O. Box 500, Batavia, IL 60510, USA}

\author{Ina Flood}
\email{iflood@umd.edu}
\affiliation{Maryland Center for Fundamental Physics, Department of Physics, University of Maryland, College Park, MD 20742, U.S.A.}

\author{Anson Hook}
\email{hook@umd.edu}
\affiliation{Maryland Center for Fundamental Physics, Department of Physics, University of Maryland, College Park, MD 20742, U.S.A.}

%%%%%%%%%%%%%%%%%%%%%%%%%%%%%%%%%%%%%%%%%%%%%%%%%%%%%%%

\vspace*{1cm}

\begin{abstract}

 We consider the general class of theories in which there is a new 
ultralight scalar field that mediates an equivalence principle 
violating, long-range force. In such a framework, the sun and the earth 
act as sources of the scalar field, leading to potentially observable 
location dependent effects on atomic and nuclear spectra. We determine 
the sensitivity of current and next-generation atomic and nuclear clocks 
to these effects and compare the results against the existing laboratory 
and astrophysical constraints on equivalence principle violating fifth 
forces. We show that in the future, the annual modulation in the 
frequencies of atomic and nuclear clocks in the laboratory caused by the 
eccentricity of the earth's orbit around the sun may offer the most 
sensitive probe of this general class of equivalence principle violating 
theories. Even greater sensitivity can be obtained by placing a 
precision clock in an eccentric orbit around the earth and searching for 
time variation in the frequency, as is done in anomalous redshift 
experiments. In particular, an anomalous redshift experiment based on 
current clock technology would already have a sensitivity to fifth 
forces that couple primarily to electrons at about the same level as 
the existing limits. Our study provides well-defined sensitivity targets 
to aim for when designing future versions of these experiments.

\end{abstract}

\maketitle

\section{Introduction}

 There are four known fundamental interactions in nature, namely 
gravity, electromagnetism, and the strong and weak nuclear forces. Taken 
together, these four forces provide an excellent fit to current 
experimental data. However, this need not be the complete picture and 
there may be additional interactions that have not yet been discovered, 
either because they are too weak or because their range is too short. 
The existence of such a fifth force constitutes one of the most 
intriguing possibilities for new physics beyond the Standard Model 
(SM)~\cite{book,Adelberger:2003zx}.
   
String theory provides motivation for the existence of a long range 
fifth force. In this class of theories, the values of the fundamental 
constants are determined by the vacuum expectation values of scalar 
fields known as moduli. These scalar fields can have wildly varying 
masses, and some of them may be extremely light, see for 
example~\cite{Damour:1994zq,Damour:1994ya}. In general, the couplings of 
moduli to matter need not respect the equivalence principle (EP). 
Therefore, such a field can serve as the mediator of a long range EP 
violating fifth force.  Nonlinearly realized discrete symmetries can 
protect the mass of a modulus against radiative corrections even if it 
has sizable couplings to the SM 
fields~\cite{Hook:2018jle,Brzeminski:2020uhm}. Apart from their 
contributions to fifth forces, moduli are a natural candidate for 
ultralight dark 
matter~\cite{Turner:1983he,Press:1989id,Sin:1992bg}. Many 
different types of searches have been proposed for this interesting 
class of dark matter 
candidates~\cite{Khmelnitsky:2013lxt,Stadnik:2014tta,Arvanitaki:2015iga,Graham:2015ouw,Graham:2015ifn,Stadnik:2015xbn,Berlin:2016woy, 
Geraci:2016fva,Krnjaic:2017zlz,Arvanitaki:2017nhi,DeRocco:2018jwe,Geraci:2018fax,Irastorza:2018dyq,Carney:2019cio,Guo:2019ker,Grote:2019uvn,Dev:2020kgz}.

There have been numerous experimental searches for new long range forces 
that violate the EP. Direct searches are based on comparing the motions 
of two bodies of different compositions in the gravitational field of a 
third. This class of searches includes experiments performed on 
suspended masses in the 
laboratory~\cite{Braginskii:1971tn,Schlamminger:2007ht,Adelberger:2009zz,Wagner:2012ui}. 
It also includes observations of the motion of free-falling objects, 
such as test masses in the field of the earth~\cite{Touboul:2017grn}, 
the moon and earth in the gravitational field of the 
sun~\cite{Dickey:1994zz,Williams:2004qba}, and gravitationally bound 
systems composed of three celestial 
bodies~\cite{Ransom:2014xla,Archibald:2018oxs}. Searches have also been 
performed that use atom interferometry to compare the rates at which 
atoms of different materials fall in the earth's gravitational 
field~\cite{Fray:2004zs,Zhou:2015pna,Zhou:2019byc,Schlippert:2014xla,Tarallo:2014oaa}. 
Although the limits from atom interferometry are not yet competitive 
with the results from experiments performed on macroscopic masses, major 
improvements are expected in the 
future~\cite{Safronova:2017xyt,MAGIS-100:2021etm}. A broad review of 
precision tests of the EP, with many additional references, may be found 
in Ref.~\cite{Tino:2020nla}.

In this paper we explore a different approach to detecting long-range 
forces that violate the EP, based on the rapidly-improving sensitivity 
of atomic and nuclear 
clocks~\cite{RevModPhys.87.637,McGrew:2018mqk,PhysRevLett.120.103201,Bothwell_2019}.  
Using precision clock experiments to search for new physics is a rapidly 
growing field, see Ref.~\cite{Safronova:2017xyt} for a review.  We 
limit our attention to the case when the fifth force is mediated by an 
ultralight scalar field. In general, the sun and the earth act as 
sources for any such scalar field. Then, since the values of fundamental 
parameters such as the fine structure constant $\alpha$ depend on the 
value of the scalar field, there are corrections to atomic and nuclear 
spectra that depend on the distance from these 
sources~\cite{Flambaum:2007ar,Shaw:2007ju,Barrow:2008se}. Atomic and 
nuclear clocks are sensitive to the frequencies of these transitions, 
and can therefore be used to search for position dependence of 
fundamental parameters. Since this effect is associated with EP 
violation~\cite{Dent:2008gu,Uzan:2010pm}, this offers an alternative 
method of searching for EP violating fifth forces. Clock experiments 
also offer a new approach to detecting more exotic fifth forces such as 
chameleon 
models~\cite{Khoury:2003aq,Khoury:2003rn,Feldman:2006wg,Mota:2006ed} 
that is distinct from the existing search methods 
\cite{Brax:2018iyo,Sakstein:2018fwz,Blinov:2018vgc}.

Clock searches for EP violation are based on comparing two atomic or 
nuclear transition frequencies against each other. These frequencies 
could be those of two different clocks at the same physical location, or 
alternatively, two clocks at separate locations. In the case of
two clocks at the same location,
as long as their transition
frequencies scale differently with fundamental parameters such as 
$\alpha$, the ratio of their frequencies will change as the distance 
from the source changes. For example, since the orbit of the earth 
around the sun is not a perfect circle, this results in an annual 
modulation in the frequencies of atomic and nuclear 
clocks~\cite{Flambaum:2007ar,Shaw:2007ju,Barrow:2008se}. We determine 
the sensitivities of current and next-generation atomic and nuclear 
clocks to this effect and compare the results against the existing 
laboratory and astrophysical constraints on EP violating fifth forces.  
We show that in the future, the annual modulation in the frequencies of 
atomic and nuclear clocks in the laboratory caused by the eccentricity 
of the earth's orbit around the sun may offer the most sensitive probe 
of this general class of EP violating theories.

Even greater sensitivity can be obtained by comparing clocks at 
different locations.  Comparing the frequency difference between a 
precision clock placed on a satellite in an eccentric orbit around the 
earth and a similar clock on earth would offer an extremely sensitive 
probe of this class of models. Experiments of this type have already 
been performed to test the general relativistic prediction for the 
gravitational redshift~\cite{Gravity_Probe_A,Delva:2018ilu}, and new 
ones proposed~\cite{Derevianko:2021kye,Tsai:2021lly,Schkolnik:2022utn}. 
Importantly, we find that such an experiment that employs current clock 
technology would already have a sensitivity to fifth forces that couple 
primarily to electrons at the about the same level as the existing 
limits. The reason for this is that direct fifth force searches are 
inherently less sensitive to forces acting on electrons, since electrons 
comprise less than 0.1\% of the mass of an atom. In contrast, atomic 
transition frequencies are extremely sensitive to the properties of the 
electron. Our analysis provides well-defined sensitivity targets to aim 
for when designing future versions of these experiments.

The outline of this paper is as follows. In the next section we consider 
the interactions of an ultralight scalar with the SM and discuss the 
current constraints on this framework from direct searches for EP 
violation and from recasting existing clock experiments. In Section III 
we study the sensitivity of next-generation atomic and nuclear clocks to 
this class of models, and show that they can explore new parameter 
space. We conclude in Section IV.

\section{Ultralight Scalar Fields and Equivalence Principle Violation}

In this section, we present a general framework for studying the effects 
of an ultralight scalar coupled to the SM. We show how direct fifth force, 
gravitational redshift, and differential redshift measurements can be 
used to place bounds on the parameters in the Lagrangian. This allows a 
concrete comparison of the sensitivities of these different experiments.

 Consider an ultralight light scalar field $\phi$ that couples to the 
particles in the SM. At energies well below the weak scale, the 
interactions of $\phi$ with the stable matter fields and the light force 
carriers of the SM can be conveniently parametrized as 
\cite{Damour2010_convention,Damour:2010rm}
 \begin{multline}
 \label{dilaton_couplings}
\mathcal{L} \supset \kappa \phi\Big[ \frac{d_e }{4 e^2} F_{\mu\nu}F^{\mu\nu}-\frac{d_g\beta_3}{2g_3}G^A_{\mu\nu}G^{A \mu\nu}-d_{m_e}m_e \bar{\psi}_e\psi_e \\-\sum_{i=u,d}{(d_{m_i}+\gamma_{m_i} d_g)m
_i\bar{\psi}_i \psi_i} \Big] \;.
 \end{multline}
 Here the parameter $\kappa$ is defined as $\kappa = \sqrt{4\pi G}$, 
where $G$ is Newton's constant. This parametrization allows a 
straightforward comparison between the force mediated by the ultralight 
scalar and gravitational effects. In this expression $e$ represents the 
charge of the electron, $g_3$ the coupling constant of quantum 
chromodynamics (QCD) and $\beta_3 \equiv \partial g_3 / \partial \log 
\mu$ its beta function. The parameters $m_i$ denote the masses of the 
fermions and $\gamma_{m} \equiv -\partial \log m / \partial \log \mu$. 
The parameters $d_x$ with $x \in \{e,g,m_i\}$ represent the couplings of 
the scalar to the corresponding gauge bosons and fermions. The couplings 
of the scalar have been parametrized such that the limit $d_g = d_{m_e} 
= d_{m_i}$ with $d_e = 0$ corresponds to the interactions of a dilaton 
that respects the equivalence principle. At these energies the SM has an 
approximate parity symmetry under which $\phi$ has been taken to be 
even. This represents the most general form of the interaction 
consistent with the symmetries up to terms of dimension 5 and linear 
order in $\phi$. In order to isolate the EP violating effects, it is 
conventional to parameterize the Lagrangian in 
Eq.~(\ref{dilaton_couplings}) in terms of the average light quark mass 
$\hat{m}\equiv(m_u+m_d)/2$ and the mass difference $\delta m \equiv 
(m_d-m_u)$ rather than the light quark masses $m_u$ and $m_d$. The 
corresponding couplings of the modulus take the form,
 \begin{equation}
\mathcal{L} \supset -\kappa \phi\left[d_{\hat{m}} \hat{m}(\bar{d} d+\bar{u} u)+\frac{d_{\delta m}}{2} \delta m(\bar{d} d-\bar{u} u)\right],
\end{equation}
where 
 \begin{align}
 \notag
  d_{\hat{m}} &= \frac{d_{m_d} m_d + d_{m_u} m_u}{m_d + m_u}\\
 d_{\delta m} &= \frac{d_{m_d} m_d - d_{m_u} m_u}{m_d - m_u}.
 \end{align}
 As a result of the interactions in Eq.~(\ref{dilaton_couplings}), the 
scalar field gives rise to a force between any two macroscopic 
bodies~\cite{Damour:2010rm}. By convention, this new force is 
parametrized in terms of its strength relative to the gravitational 
force. Accordingly, the potential energy $V$, which includes the effects 
of both the gravitational force and the new force, now takes the form,
 \bea 
 \label{eq: gravitational potential} 
V = -G\frac{m_{\bf A} m_{\bf B}}{r_{\bf AB}}(1+\alpha_{\bf A} 
\alpha_{\bf B}e^{-\frac{r_{\bf AB}}{\lambda}}) \;. 
 \eea 
 Here $m_{\bf A}$ and $m_{\bf B}$ are the masses of the two bodies {\bf 
A} and {\bf B}, $r_{\bf AB}$ is the distance between them, and $\lambda 
\equiv 1/m_\phi$ sets the range of the interaction. The parameters 
$\alpha_{\bf A}$ and $\alpha_{\bf B}$, which are functions of $d_e$, 
$d_g$, $d_{m_e}$ and $d_{m_i}$, depend on the compositions of {\bf A} 
and {\bf B}. Therefore, in general, the force mediated by $\phi$ 
violates the EP. The parameters $\alpha_{\bf A}$ and 
$\alpha_{\bf B}$ can be approximated as,
 \bea \label{eq: alpha_x}
\alpha_{\bf X} \simeq 
d_{g}^{*} 
+ &\left[ ( d_{\hat{m}}-d_{g} ) Q_{\hat{m}} + (d_{\delta m}-d_{g})Q_{\delta m} +(d_{m_e}-d_{g})Q_{m_e} \right.\\
 &+ \left. d_{e} Q_{e} \right]_{\bf X} \; ,
 \eea
where the composition-independent part $d_{g}^{*}$ is given by 
 \begin{equation}
     d_{g}^{*} \equiv 
d_{g} + 0.093(d_{\hat{m}}-d_{g})+ 10^{-4}[2.7 d_e + 2.75  (d_{m_e}-d_{g})].
 \end{equation}
 The remaining composition-dependent part of $\alpha_{\mathbf{X}}$ in 
Eq.~(\ref{eq: alpha_x}) is parameterized in terms of the variables 
$Q_{\hat{m}}$, $Q_{\delta m}$, $Q_{e}$ and $Q_{m_e}$ that depend on the 
mass number $A$ and atomic number $Z$ of the atomic nuclei of which the 
body is composed{\footnote{Our parametrization of $\alpha_{\bf X}$ 
differs from that in Ref.~\cite{Damour:2010rm}, and so our expressions 
for $Q_{\hat{m}}$, $Q_{\delta m}$, $Q_{e}$ and $Q_{m_e}$ are also 
different.}},
  \bea\label{eq: alpha_x_defs}
Q_{\hat{m}} &\equiv
-\frac{0.036}{A^{1/3}}-1.4 \times 10^{-4} \frac{Z(Z-1)}{A^{4/3}} - 0.02 \frac{(A-2Z)^2}{A^2}, \\ 
Q_{\delta m} &\equiv 1.7 \times 10^{-3} \frac{A-2Z}{A}, \\
Q_{e} &\equiv
7.7 \times 10^{-4}\frac{Z(Z-1)}{A^{4/3}} + 8.2 \times 10^{-4} \left(\frac{Z}{A}-\frac{1}{2}\right), \\
Q_{m_e} &\equiv 5.5 \times 10^{-4} \left(\frac{Z}{A}-\frac{1}{2}\right).
 \eea 
 We see from this 
that $\alpha_{\bf X}$ naturally splits up into a composition-independent 
term $d_{g}^{*}$ and a composition-dependent term that is contained in 
the square bracket in Eq.~(\ref{eq: alpha_x}). For a general choice of 
modulus couplings the term in the square brackets does not vanish, so 
the contribution to the potential from the ultralight scalar depends on 
the compositions of the test bodies. Therefore the resulting force 
violates the weak EP. The existing limits on EP violating forces can 
be translated into bounds on the couplings of the scalar $\phi$. 

We see from Eq.~\eqref{eq: alpha_x_defs} that $|Q_{\delta m}| \ll
|Q_{\hat{m}}|$. We therefore expect that, in general, experiments 
searching for EP violation will be much more sensitive to the coupling 
$d_{\hat{m}}$ than to $d_{\delta m}$. For simplicity, we will therefore 
neglect the coupling $d_{\delta m}$ in the discussion that follows.

Apart from generating an EP violating force, the interactions in 
Eq.~(\ref{dilaton_couplings}) imply that the effective values of 
fundamental constants such as $\alpha$ and $m_e$ at any given location 
depend on the value of $\phi$ at that location. For example, for these 
two parameters we have,
 \begin{equation} \label{eq: alpha spatial variation}
\alpha(x) = \bar{\alpha}[1 + d_e \kappa \phi(x)] \quad m_e(x) = \bar{m_e}[1 + d_{m_e} \kappa \phi(x)] \; .
 \end{equation}

 Here $\bar{\alpha}$ ($\bar{m_e}$) denotes the value of the fine 
structure constant (electron mass) in the absence of the terms in 
Eq.~(\ref{dilaton_couplings}). Consequently the sourcing of $\phi$ by 
massive objects such as the sun and the earth causes the value of 
fundamental constants such as $\alpha$ and $m_e$ to depend on the 
distance from these sources. This offers an alternative method of 
probing this class of models using atomic and nuclear clocks, which is 
the focus of this paper.

\subsection{Direct Fifth Force Measurements} \label{sec: Direct fifth force}

In this subsection we review how to map the limits from direct fifth 
force searches onto bounds on the parameters in 
Eq.~\eqref{dilaton_couplings}.  The results are summarized in 
Eqs.~\eqref{eq: dg massless}, \eqref{eq: de massless}, \eqref{eq: dm 
massless} and \eqref{eq: dme massless}.

At present, the most precise tests of EP violation are based on 
measurements of how two test bodies {\bf A} and {\bf B} composed of 
different materials accelerate towards a third body {\bf C}, which is 
usually the earth or the sun. If the EP holds, the two accelerations 
should be identical. By convention, the experimental limits on EP 
violation are expressed in terms of the Eotvos parameter,
 \bea
\label{eq: eotvos}
    \eta \equiv 2 \frac{\vert \vec{a}_{\bf A}-\vec{a}_{\bf B}\vert}{\vert \vec{a}_{\bf A}+\vec{a}_{\bf B} \vert}
  \eea
 where $a_{\bf A}$ and $a_{\bf B}$ represent the accelerations of the 
test bodies {\bf A} and {\bf B}. Since the interactions in 
Eq.~(\ref{dilaton_couplings}) give rise to an EP violating force between 
any two macroscopic bodies, the experimental limits on $\eta$ can be 
translated into bounds on the couplings of the ultralight scalar to 
matter.

\begin{figure}[t]
    \centering
    \includegraphics[width=0.995\linewidth]{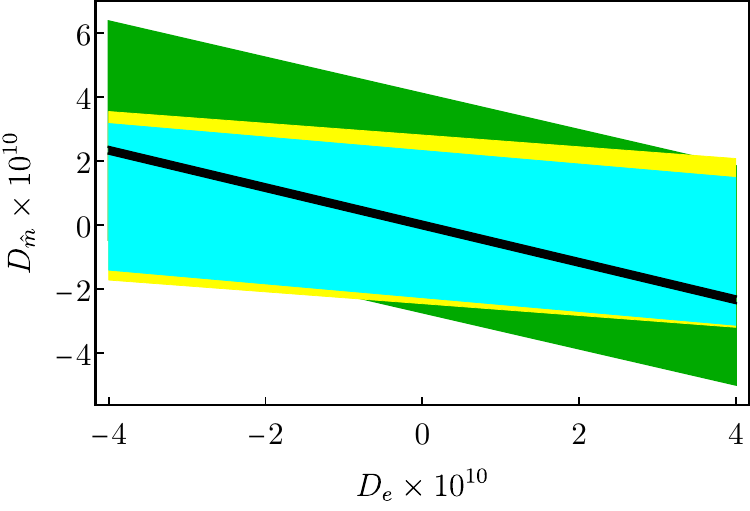}
    \caption{Current bounds on $D_{e}$ vs. $D_{\hat{m}}$ set by fifth force 
experiments 
\cite{Wagner:2012ui,Touboul:2017grn,Berge2017_MICROSCOPE,Braginskii:1971tn}. 
The black band represents the bound set by MICROSCOPE 
\cite{Berge2017_MICROSCOPE, Touboul:2017grn}, the blue and yellow bands 
are the constraints set by the EotWash group with Be-Ti and Be-Al masses 
\cite{Wagner:2012ui}, and the green band is the Moscow group's result 
obtained with Al-Pt masses \cite{Braginskii:1971tn}. }
    \label{fig: EP experiments}
 \end{figure}

 In the limit that the distances between the bodies {\bf A},{\bf B} 
and {\bf C} are all much smaller than $\lambda$, so that the mass of the 
modulus can be neglected, we can estimate the Eotvos parameter in this 
class of models from Eq.~(\ref{eq: gravitational potential}) as,
 \bea 
\label{eq: eotvos_dilaton}
    \eta &\approx (\alpha_{\bf A}-\alpha_{\bf B})\alpha_{\bf C} \\
    &\approx \left[\Delta Q_{\hat{m}} (d_{\hat{m}}-d_{g}) + \Delta Q_{e} d_{e} + \Delta Q_{m_e} (d_{m_e} - d_g)\right] 
\alpha_{\bf C} \; .
  \eea
In most simple models the composition independent part of $\alpha_{\bf C}$ 
will dominate over composition dependent part. In this case we can make 
the approximation $\alpha_{\bf C} \approx d_{g}^{*}$ so that
 \bea
\eta &\approx 
\left[\Delta Q_{\hat{m}} (d_{\hat{m}}-d_{g}) + \Delta Q_{e} d_{e} + \Delta Q_{m_e} (d_{m_e} - d_g) \right] d_{g}^{*} \\
 &\approx \Delta Q_{\hat{m}} D_{\hat m} + \Delta Q_{e} D_{e} + \Delta Q_{m_e} D_{m_e} \;. 
 \eea
 Here we have defined $D_{e} \equiv d^{*}_{g} d_{e}$, $D_{\hat{m}} 
\equiv d^{*}_{g} (d_{\hat{m}}-d_{g})$ and $D_{m_e} \equiv d^{*}_{g} 
(d_{m_e}-d_{g})$.  Since typically $\Delta Q_{m_e} \ll \Delta Q_{e} \, , 
\Delta Q_{\hat{m}} $, using the experimental bound on $\eta$ for two 
given test bodies of known compositions, the allowed 
$(D_{\hat{m}},D_{e})$ parameter space can be constrained to a band as 
shown in Fig. \ref{fig: EP experiments}. The most accurate measurement 
of $\eta$, performed by the MICROSCOPE mission~\cite{Touboul:2017grn, 
Berge2017_MICROSCOPE}, sets constraints at the $10^{-14}$ level and is 
represented by the black band in Fig. \ref{fig: EP experiments}.

 In addition to constraining the $D_{\hat{m}}$ and $D_{e}$ parameters, 
we can translate the bounds on the Eotvos parameter $\eta$ into limits 
on the individual modulus couplings $d_{x}$. For example, using Eq. 
\eqref{eq: eotvos_dilaton} we can set an upper bound on $d_g$ assuming that all 
the other couplings vanish,
 \begin{equation} \label{eq: dg massless}
    \eta \approx \Delta Q_{\hat{m}} d_{g}^2  \; \; \;  \Rightarrow \; \; \; d_g = \sqrt{\frac{\eta}{\Delta Q_{\hat{m}}}} \; , d_{\hat{m}} = d_{m_e} = d_e = 0.
 \end{equation}
 
 The analogous bounds on $d_e$, $d_{\hat{m}}$ and $d_{m_e}$ are given 
by,
 \begin{equation} \label{eq: de massless}
 d_e = \sqrt{\frac{\eta}{\Delta Q_{e}(2.7 \times 10^{-4}+ Q_{e,\bf C})}} \; , \; d_{\hat{m}} = d_{m_e} = d_g = 0,
 \end{equation}
 \begin{equation} \label{eq: dm massless}
 d_{\hat{m}} = \sqrt{\frac{\eta}{\Delta Q_{\hat{m}}(9.3 \times 10^{-2}+ Q_{\hat{m},\bf C})}} \; , \; d_{m_e} = d_e = d_g = 0,
 \end{equation}
  \begin{equation} \label{eq: dme massless}
 d_{m_e} = \sqrt{\frac{\eta}{\Delta Q_{m_e}(2.75 \times 10^{-4}+ Q_{m_e,\bf C})}} \; , \; d_{\hat{m}} = d_e = d_g = 0.
 \end{equation}
  By performing multiple measurements on bodies of different 
compositions, it is in principle possible to set independent constraints 
on all of the $D_x$ and $d_x$ parameters.

Although these bounds have been obtained under the assumption that the 
distances between the bodies {\bf A}, {\bf B} and {\bf C} are all much 
smaller than $\lambda$, the extension of these limits to the more 
general case is straightforward~\cite{Adelberger:2003zx, 
Berge2017_MICROSCOPE},
 \begin{equation} \label{eq: dx massive}
    d_x = \frac{d_x^{\text{massless}}}{\sqrt{\Phi (\frac{R_{\bf C}}{\lambda})(1+\frac{r}{\lambda})e^{-\frac{r}{\lambda}}}} \;.
 \end{equation}
 Here $\Phi(x) \equiv 3(x \cosh{x}-\sinh{x})/x^3 $, $R_{\bf C}$ is the radius 
of the body {\bf C}, and $d_x^{\text{massless}}$ represents the 
corresponding bound in the limit that $m_{\phi}=0$, given by 
Eqs.~\eqref{eq: dg massless}, \eqref{eq: de massless}, \eqref{eq: dm 
massless} and \eqref{eq: dme massless}.

\subsection{Clock Experiments} \label{sec: clock experiments}

In this subsection, we show how to map the results of clock experiments 
onto the parameter space shown in Eq.~\eqref{dilaton_couplings}.  This 
allows for a direct comparison between clock experiments and direct 
fifth force measurements.

While direct measurement of the fifth force currently provides the most 
stringent constraints on the couplings $d_x$ of the ultralight scalar, 
there is an alternative method to constrain the new force. Recall that the 
couplings introduced in Eq.~\eqref{dilaton_couplings} modify the 
potential energy between two masses as shown in Eq.~\eqref{eq: gravitational potential}. The correction to the potential energy can be 
rewritten in the familiar form,
 \bea \label{eq: yukawa potential energy}
\delta V = -G\frac{m_{\bf A} m_{\bf B}}{r_{\bf AB}}\alpha_{\bf A} 
\alpha_{\bf B}e^{-\frac{r_{\bf AB}}{\lambda}} = - \frac{q_{\bf A} q_{\bf B}}{4 \pi r_{\bf AB}}e^{-\frac{r_{\bf AB}}{\lambda}} \; , 
 \eea
 where $q_{\bf X} = \kappa \alpha_{\bf X} m_{\bf X}$ represents the 
charge of the body under the new Yukawa force. From the above expression 
we see that each massive body {\bf X} sources the scalar field $\phi$ 
as,
 \bea \label{eq: yukawa potential}
\phi_{\bf X} = - \frac{q_{\bf X}}{4 \pi r}e^{-\frac{r}{\lambda}}. 
 \eea
 As can be seen from Eq.~\eqref{eq: alpha spatial variation}, the scalar 
field $\phi$ affects the values of fundamental constants. This results 
in a spatial variation in the values of fundamental constants in the 
vicinity of a source body {\bf X}. For example, in the case of the fine 
structure constant we have,
 \bea \label{eq: alpha gravitational potential}
\frac{\Delta \alpha}{\alpha} = d_e \kappa \phi_{\bf X} = -\frac{d_e \alpha_{\bf X} G m_{\bf X}}{r}e^{-\frac{r}{\lambda}} =  d_e \alpha_{\bf X} U_{\bf X} e^{-\frac{r}{\lambda}} \;.
 \eea
 Here $U_{\bf X}$ represents the gravitational potential sourced by the 
body {\bf X}. From the above equation, we see that measuring the 
variation of fundamental constants in the neighborhood of a massive body 
such as the earth or sun allows us to probe the same couplings that 
direct fifth force searches are sensitive to. We now discuss the 
sensitivity of atomic and nuclear clocks to this variation.

\begin{figure*}[t] 
\centering
\begin{minipage}{.48\textwidth}
    \centering
    \includegraphics[width=0.995\linewidth]{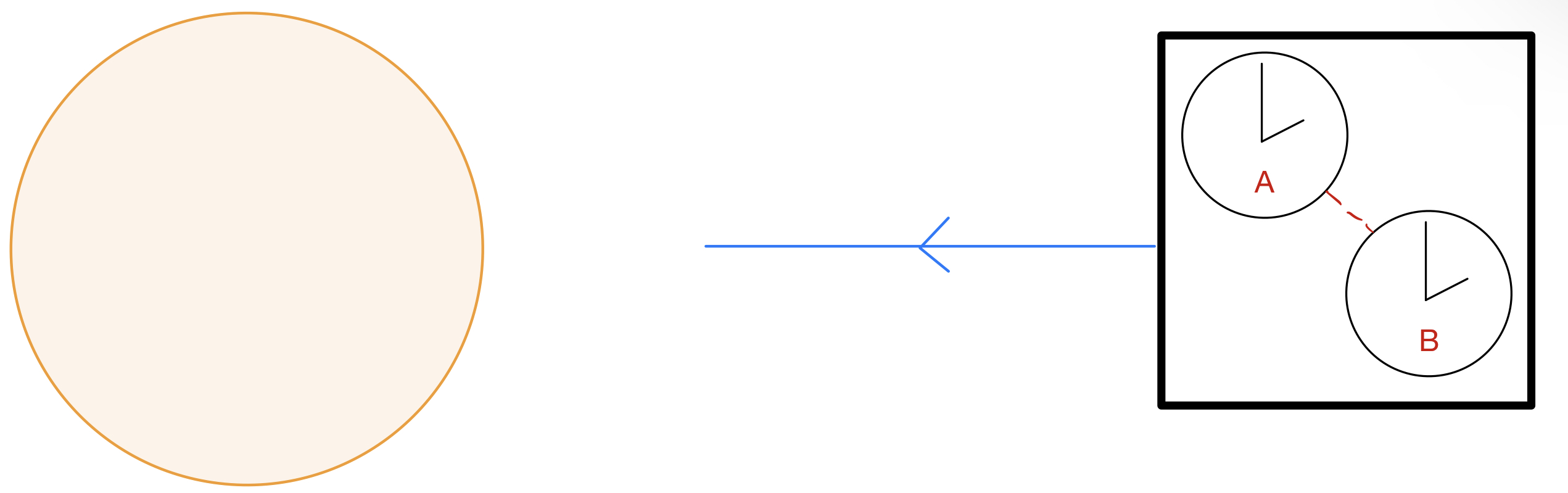}
    \caption{A schematic picture of an experiment involving two 
different clocks that travel together and experience 
the same change in the gravitational potential.  The location dependence 
of the ratio of the two transition frequencies is a sensitive probe of 
EP violation. }
    \label{fig: clocks 2}
\end{minipage}%
\hfill
\begin{minipage}{.48\textwidth}
    \centering
    \includegraphics[width=0.995\linewidth]{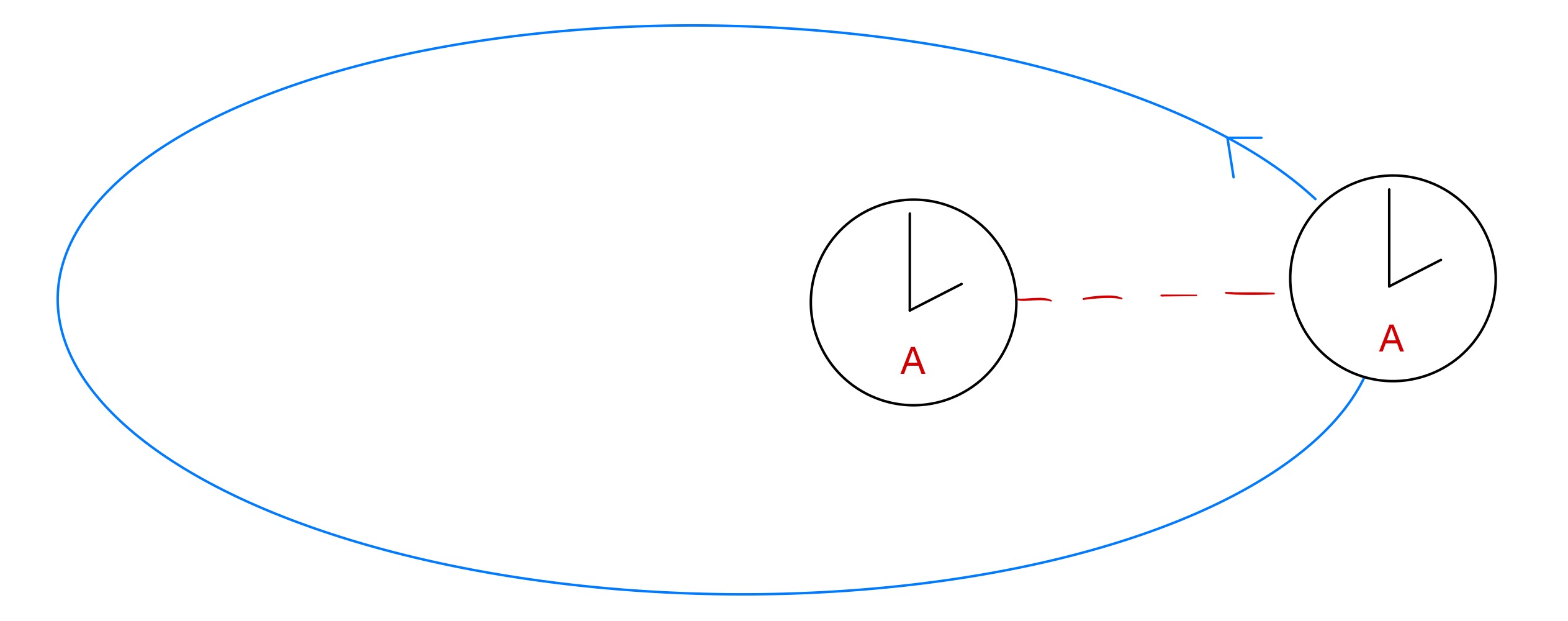}
    \caption{A schematic picture of an experiment involving two 
identical clocks that are spatially separated. The dependence of the 
difference in frequencies between the two clocks on the difference in 
the gravitational potential between their locations is a sensitive probe 
of EP violation.}
    \label{fig: clocks 1}
\end{minipage}
\end{figure*}

The principle of Local Position Invariance (LPI) states that the frequency 
of any given clock in its local frame is independent of its position in
space. This can be expressed as
  \begin{equation} \label{eq: LPI}
     f^{local}_A(x) = f^{local}_A(\infty) ,
 \end{equation}
 where $f^{local}_A(\infty)$ denotes the frequency of the clock A at 
infinity, where the gravitational potential $U$ vanishes. It follows from 
Eq.~\eqref{eq: alpha gravitational potential} that in the class of theories 
we are considering the values of the fundamental constants are not the 
same at different points in space. Then the frequencies of clocks depend 
on their location in space and so the principle of LPI is no longer 
valid. Therefore Eq.~(\ref{eq: LPI}) will receive corrections whose 
magnitude will, in general, depend on the type of the clock used in the 
experiment. This violation of LPI can be parametrized in terms of the 
anomalous redshift parameter $\beta_A$ as
 \begin{equation} \label{eq: anomalous redshift definition}
\frac{f^{local}_A(x) - f^{local}_A(\infty) }{f^{local}_A(\infty)} =
\beta_A U(x) \; ,
 \end{equation}
 where values of $\beta_A$ different from zero are the result of new 
physics. The subscript $A$ indicates the type of clock being employed. 
In practice, when comparing two clocks, one typically measures 
frequencies in the rest frame of one of the clocks, which we identify 
with the lab frame. This leads to the relation~\footnote{
 In the literature, the anomalous 
redshift parameter $\beta$ is conventionally defined as $\frac{f^{\infty}_A(x) - 
f^{\infty}_A(\infty)}{f^{\infty}_A(\infty)} = (1+ \beta )U(x)$, as seen 
by the observer at infinity. This definition implicitly assumes that the 
anomalous scaling is independent of the clock used and is therefore not 
suitable for our purposes.
 }
 \begin{equation} 
 \label{eq: anomalous redshift lab frame}
 \frac{f^{\text{lab}}_A(x) - f^{\text{lab}}_A(x_{\text{lab}})}{f^{\text{lab}}_A(x_{\text{lab}})} =  
(1+\beta_A) [U(x)-U(x_{\text{lab}})] \; .
 \end{equation}
 Here the term $1$ in the bracket represents the standard prediction from 
general relativity and $x_{\text{lab}}$ denotes the position of the lab.

The dependence of a clock transition on fundamental parameters is 
conventionally expressed as
 \begin{equation} \label{eq: frequency scaling}
    f_A \propto R \, \alpha^{K^{A}_{\alpha}} \mu^{K^{A}_{\mu}} X_{q}^{K^{A}_{q}} \propto m_e \, \alpha^{K^{A}_{\alpha}+2} \mu^{K^{A}_{\mu}} X_{q}^{K^{A}_{q}},
\end{equation}
 where $\alpha$ is the fine structure constant, $\mu \equiv 
{m_{p}}/{m_{e}}$ is the proton-to-electron mass ratio, $X_q \equiv 
{m_{q}}/{\Lambda_{QCD}}$ 
is the ratio of the average light quark mass to the QCD 
scale and $R \propto m_e \alpha^2$ denotes the Rydberg constant. The 
coefficients $K^{A}_{\alpha,\mu,q}$ characterize the sensitivity of a 
given transition to variations of the corresponding parameters.  
Typically, $K^{A}_{\mu} = -1$ for hyperfine transitions, $K^{A}_{\mu} 
= 0$ for optical and $K^{A}_{\mu} 
= 1$ for nuclear transitions. The $K^{A}_{\alpha,q}$ have to 
be determined numerically for each transition.
  
From Eqs. \eqref{eq: anomalous redshift definition} and \eqref{eq: 
frequency scaling} we find 
 \bea 
 \label{eq: clock comparison constants} 
\beta_{A} U(x) 
= \frac{\Delta m_{e}}{m_{e}} 
+ (K^{A}_{\alpha}+2) \frac{\Delta \alpha}{\alpha} 
+ K^{A}_{\mu} \frac{\Delta \mu}{\mu} + 
K^{A}_{q} \frac{\Delta X_{q}}{X_{q}} \; ,
 \eea 
 where $\frac{\Delta X}{X} = \frac{X(x)-X(\infty)}{X(\infty)}$ with $X 
\in \{ \alpha, \mu, X_{q}, m_{e} \}$. Using this equation, constraints 
on $\beta_A$ can be turned into bounds on the variation of various 
fundamental constants.

In order to compare these bounds with those from fifth force experiments 
we need to express the variation of the fundamental parameters in terms of 
the couplings shown in Eq. \eqref{dilaton_couplings}\footnote{
 The following discussion implicitly assumes that the gravitational potential can 
be independently determined in the presence of the fifth force.  We 
study this potential complication in Appendix~\ref{appendix: deltaU} and 
show that it has only a small effect on the results of this section.}.
 A brief calculation yields
 \bea \label{eq: variation of constants}
\frac{\Delta \alpha}{\alpha} &=  d_{e} \alpha_{\bf X}  U \simeq  D_{e}  U, \\
\frac{\Delta \mu}{\mu} &= -(d_{m_{e}}-d_{g}) \alpha_{\bf X}  U \simeq -D_{m_{e}}  U, \\
\frac{\Delta X_{q}}{X_{q}} &= (d_{\hat{m}}-d_{g}) \alpha_{\bf X}  U \simeq D_{\hat{m}}  U, \\
\frac{\Delta m_{e}}{m_{e}} &= d_{m_{e}} \alpha_{\bf X}  U \simeq (D_{m_{e}}+D_g)  U \;,
 \eea
 where in the second step we have approximated $\alpha_{\bf X} \approx 
d_g^*$ and defined $D_g \equiv d_g d_g^*$. These expressions are valid 
for $r\ll \lambda$. Experiments that constrain the variation of 
fundamental constants sometimes employ the alternative parameterization, 
${\Delta X}/{X} = k_X U$. From Eq.~\eqref{eq: variation of constants}, 
we see that $k_{\alpha} = D_e$, $k_{\mu} = - D_{m_{e}}$, and 
$k_{q}=D_{\hat{m}}$. The translation of bounds between the two different 
parametrizations is therefore straightforward.

From Eqs. \eqref{eq: clock comparison constants} and \eqref{eq: 
variation of constants} we can express $\beta_A$ in terms of the 
couplings of the modulus,
 \bea \label{eq: beta A}
\beta_{A} &= \left[(K^{A}_{\alpha}+2) d_{e} - K^{A}_{\mu} (d_{m_{e}}-d_g) + K^{A}_{q} (d_{\hat{m}}-d_g) + d_{m_{e}} \right]\alpha_{\bf X} \\
&\approx (K^{A}_{\alpha}+2) D_{e} + (1-K^{A}_{\mu}) D_{m_{e}} + K^{A}_{q} D_{\hat{m}} + D_g \;.
 \eea
 Since direct fifth force measurements are also sensitive to the 
$D_{x}$, the relation above allows for a direct comparison between these 
experiments and clock experiments. 

 We can translate the bounds on $\beta_A$ into bounds on the modulus 
couplings $d_x$ under the assumption that only one coupling has a 
non-zero value,
 \begin{equation} \label{eq: dg massless clocks}
 d_g = \sqrt{\frac{\beta_A}{\vert K_{\mu}^{A}-K_{q}^{A} \vert} } \; , \; d_{\hat{m}} = d_e = d_{m_e} = 0,
 \end{equation}
 \begin{equation} \label{eq: de massless clocks}
 d_e = \sqrt{\frac{\beta_A}{\vert K_{\alpha}^{A}+2 \vert (2.7 \times 10^{-4}+ Q_{e, X})}} \; , \; d_{\hat{m}} = d_g = d_{m_e} = 0,
 \end{equation}
 \begin{equation} \label{eq: dm massless clocks}
 d_{\hat{m}} = \sqrt{\frac{\beta_A}{\vert K_{q}^{A} \vert (9.3 \times 10^{-2}+ Q_{\hat{m},X})}} \; , \; d_e = d_g = d_{m_e} = 0.
 \end{equation}
  \begin{equation} \label{eq: dme massless clocks}
 d_{m_e} = \sqrt{\frac{\beta_A}{\vert 1-K_{\mu}^{A}\vert \, (2.75 \times 10^{-4} + Q_{m_{e},X})}} \; , \; d_e = d_g = d_{\hat{m}} = 0.
 \end{equation}
 By employing multiple clocks, each of a different composition, it is in 
principle possible to set independent constraints on all of the $D_x$ 
and $d_x$ parameters.

Our results have been derived under the assumption that the experiments were 
performed at distances such that $r\ll \lambda$. When we relax this
assumption, the above formulas generalize to \cite{Adelberger:2003zx}
 \begin{equation} \label{eq: dx massive clocks}
    d_x = \frac{d_x^{\text{massless}}}{\sqrt{\Phi (\frac{R_{\bm X}}{\lambda})e^{-\frac{r}{\lambda}}}},
 \end{equation}
 where $d_x^{\text{massless}}$ is the corresponding parameter given in
Eqs.~\eqref{eq: dg massless clocks}, \eqref{eq: de massless clocks}, 
\eqref{eq: dm massless clocks} and \eqref{eq: dme massless clocks}.

The clock experiments that search for fifth forces fall into two 
distinct classes, differential redshift measurements and gravitational 
redshift measurements, illustrated in Figs.~\ref{fig: clocks 2} and 
\ref{fig: clocks 1} respectively.  Differential redshift measurements 
involve comparing two clocks composed of different materials at the same 
location as they orbit another body. Gravitational redshift measurements 
involve comparing the frequencies of a clock in orbit around the earth 
with a clock on earth. This class of experiments is sensitive 
to the redshift predicted by general relativity and bounds are placed on 
any additional source of redshift. We now consider the existing limits 
from these two classes of experiments in turn.
 
We first consider differential redshift measurements. A comparison of 
two different clocks $A$ and $B$ that experience the same change in the 
potential allows a measurement of the difference $(\beta_A - \beta_B)$. 
Then, using Eq. \eqref{eq: beta A}, we can easily recover information 
about the couplings of the modulus,
 \bea \label{eq: beta AB}
\beta_{AB} &= [\Delta K^{AB}_{\alpha} d_{e} - \Delta K^{AB}_{\mu} (d_{m_{e}}-d_g) + \Delta K^{AB}_{q} (d_{\hat{m}}-d_g)]\alpha_{\bf X} \\
&\approx \Delta K^{AB}_{\alpha} D_{e} - \Delta K^{AB}_{\mu} D_{m_{e}} + \Delta K^{AB}_{q} D_{\hat{m}} \;.
 \eea
 Here we have defined, $\beta_{AB} \equiv \beta_A - \beta_B$ and $\Delta K^{AB}_X \equiv K^{A}_X - K^{B}_X $.
 
 A natural realization of the experiment involves comparing the 
frequencies of two different clocks in the laboratory over the course of 
a year~\cite{PhysRevLett.111.060801,Leefer:2016xfu,Lange:2020cul}. As the distance 
between earth and the sun changes due to the eccentricity of the orbit, 
the frequencies of the clocks change accordingly. The ratio of the 
frequencies of the two clocks, $\tilde{f}(x) \equiv 
f^{local}_A(x)/f^{local}_B(x)$ changes as
 \begin{equation} 
\label{eq: differential redshift experiment}
    \frac{\tilde{f}(x(t)) - \tilde{f}(x(t_0))}{\tilde{f}(x(t_0))} = (\beta_A - \beta_B ) \left[U(x(t))-U(x(t_0))\right] \;.
 \end{equation}
 Therefore, by comparing the frequencies of the clocks over the course 
of a year, we can obtain a measurement of $(\beta_A - \beta_B)$. 
Currently, state-of-the-art experiments constrain this ratio at the 
$10^{-7}$ level \cite{Lange:2020cul}, which results in the following 
limits,
 \bea
D_e \lesssim 10^{-8}, \quad
D_{m_e} \lesssim 10^{-6},\quad
D_{\hat{m}} \lesssim  10^{-6} \; .
 \eea
 From Fig.~\ref{fig: EP experiments}, we see that the limits from
this class of atomic 
clock experiments are currently at least 3 orders of magnitude weaker than the 
direct limits from fifth force measurements in the $(D_{e},D_{\hat{m}})$ 
parameter space.

We now turn our attention to gravitational redshift measurements. As Eq. 
\eqref{eq: anomalous redshift lab frame} suggests, $\beta_A$ can be 
determined by comparing two identical spatially separated clocks, as 
illustrated in Fig.~\ref{fig: clocks 1}. This approach is also not new. 
The first successful realization of this method was achieved by Gravity 
Probe - A \cite{Gravity_Probe_A}, where the 
frequency of a microwave clock on a satellite was compared with the 
frequency of an identical clock on earth via microwave link as the 
satellite was changing altitude. After accounting for special and 
general relativistic effects, the measured frequency was transformed to 
the local frame of satellite leading to a constraint on the anomalous 
redshift via the relation,
  \begin{equation} 
 \label{eq: anomalous redshift experiment}
    \frac{f^{local}_A(x) - f^{local}_A(x_{\oplus})}{f^{local}_A(x_\oplus)} = \beta_A  \left[U(x)-U(x_{\oplus})\right],
  \end{equation}
    where $x_\oplus$ denotes the position of the clock on earth.

 This experiment constrained $\beta_A$ at the level of $10^{-4}$. 
Although this bound was later improved by an order of magnitude by the 
Galileo satellite~\cite{Delva:2018ilu}, it remains many orders of 
magnitude below the strongest limits from direct fifth force searches.

\section{Experimental prospects}

\begin{figure}[h]
    \centering
    \includegraphics[width=0.995\linewidth]{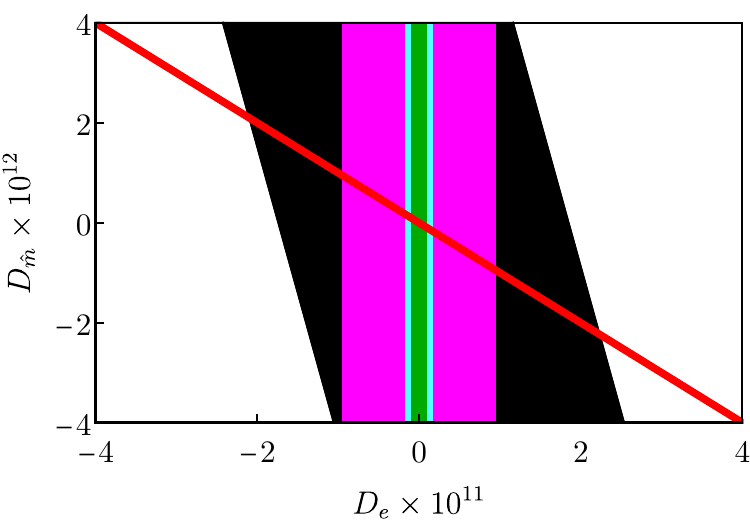}
    \caption{
    In the $D_{\hat{m}}$ vs. $D_{e}$ plane, we show how the projected limits 
from future earth and space based differential redshift experiments compare against 
the current bounds from direct fifth force searches. The black region 
represents the current bound set by MICROSCOPE 
\cite{Touboul:2017grn,Berge2017_MICROSCOPE}. The magenta and light cyan lines 
show the projected sensitivity of the SpaceQ experiment 
\cite{Tsai:2021lly} while traveling towards the $r=0.39$ AU and $r=0.1$ AU 
orbits respectively, assuming that the satellite is equipped with two 
optical clocks with $\Delta K_{\alpha} = 7$ and ${\Delta \tilde{f}}/{\tilde{f}} = 
10^{-18}$. The green band shows the projected sensitivity of an earth 
based experiment based on two optical clocks with $\Delta K_{\alpha} = 
7$, $\Delta K_{q} = 0$ and ${\Delta \tilde{f}}/{\tilde{f}} = 10^{-21}$.  
The red line shows the bound that could be set by an earth based nuclear 
clock - optical clock system, with $\Delta K_{\alpha} = 10^4$, $\Delta 
K_{q} = 10^5$ and ${\Delta \tilde{f}}/{\tilde{f}} = 10^{-18}$. }
    \label{fig: EP experiments 2}
\end{figure}

\begin{figure*}[h] 
\centering
\begin{minipage}{.48\textwidth}
    \centering
    \includegraphics[width=0.995\linewidth]{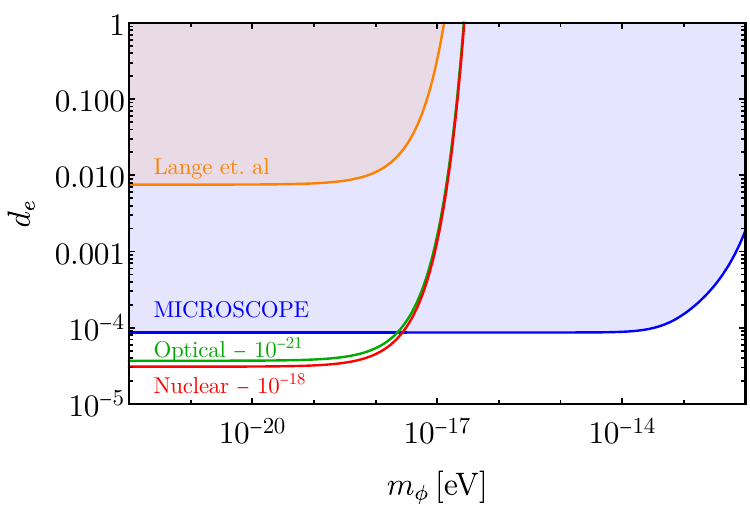}
    \caption{In the $d_e$ vs. $m_\phi$ plane, we show how the projected 
limits from future earth based two-clock experiments compare against the 
current bounds from direct fifth force searches.  The orange region 
represents the current bounds from atomic clock 
experiments~\cite{Lange:2020cul}.  The blue region shows the parameter 
space excluded by the MICROSCOPE 
experiment~\cite{Touboul:2017grn,Berge2017_MICROSCOPE}. The green line 
depicts the projected limit from a future Earth based experiment 
involving two optical clocks with $\Delta K_{\alpha} = 7$ and ${\Delta 
\tilde{f}}/{\tilde{f}} = 10^{-21}$. The red line shows the projected 
sensitivity of an Earth based two-clock experiment involving a nuclear 
clock with $\Delta K_{\alpha} = 10^4$ and ${\Delta \tilde{f}}/{\tilde{f}} 
= 10^{-18}$. \newline\newline\newline\newline\newline\newline\newline}
    \label{fig: de_bounds}
\end{minipage}%
\hfill
\begin{minipage}{.48\textwidth}
    \centering
    \includegraphics[width=0.995\linewidth]{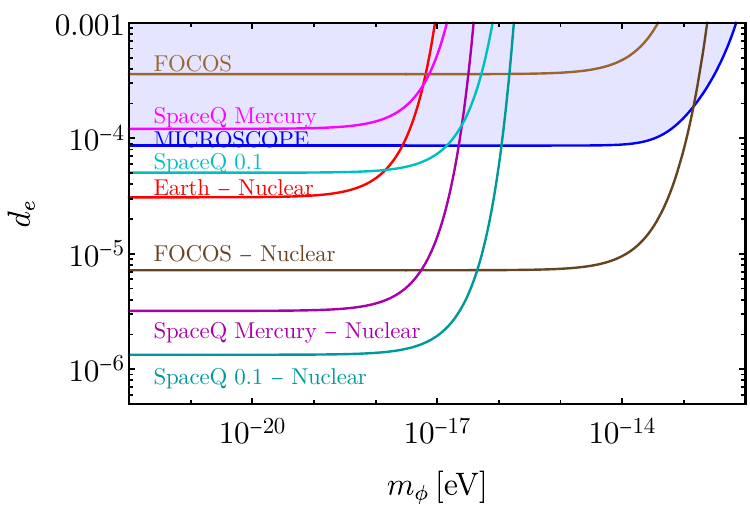}
    \caption{In the $d_e$ vs. $m_\phi$ plane we show how the projected 
limits from future experiments compare against the current bound from 
direct fifth force searches. The blue region shows the parameter space 
excluded by the MICROSCOPE 
experiment~\cite{Touboul:2017grn,Berge2017_MICROSCOPE}. The red line 
shows the sensitivity of an Earth based two-clock experiment involving a 
nuclear clock with $\Delta K_{\alpha} = 10^4$ and $\Delta \tilde{f}/\tilde{f} = 
10^{-18}$. The brown and dark brown lines show the potential reach of 
the FOCOS experiment \cite{Derevianko:2021kye}, assuming that the 
satellite is equipped with an optical clock with $K_{\alpha} = -6$ or a 
nuclear clock with $K_{\alpha} = 10^4$ respectively and measures the redshift with the accuracy of $\beta = 10^{-9}$. The magenta and 
cyan lines depict the sensitivity of the SpaceQ mission 
\cite{Tsai:2021lly} while traveling towards the $r=0.39$ AU and $r=0.1$ 
AU orbits respectively, assuming that the satellite is equipped with two 
optical clocks with $\Delta K_{\alpha} = 7$ and $\Delta \tilde{f}/\tilde{f} = 
10^{-18}$. The dark magenta and dark cyan lines depict the ultimate 
sensitivities of the SpaceQ experiment while traveling towards the 
$r=0.39$ AU and $r=0.1$ AU orbits respectively, when the satellite is 
equipped with a nuclear-optical clock system with $\Delta 
K_{\alpha} = 10^4$ and $\Delta \tilde{f}/\tilde{f}  = 10^{-18}$.
}
    \label{fig: focos}
\end{minipage}
\vskip\baselineskip
% \end{figure*}

% \begin{figure*}[h] 
% \centering
\begin{minipage}{.48\textwidth}
    \centering
    \includegraphics[width=0.995\linewidth]{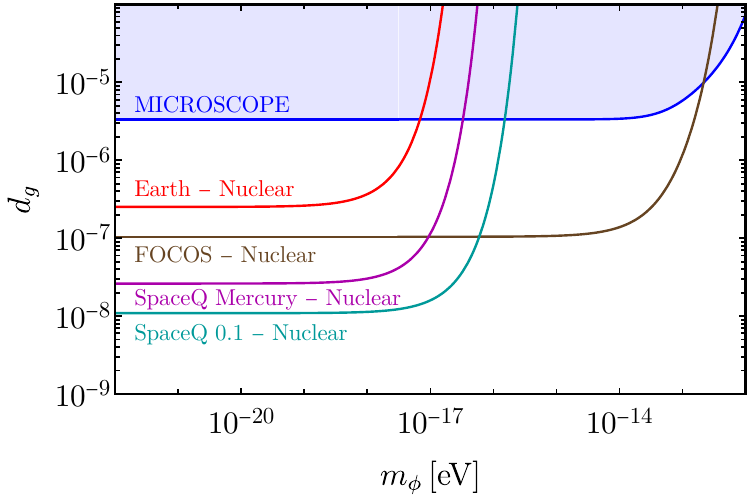}
    \caption{In the $d_g$ vs. $m_\phi$ plane we show how the projected 
limits from future clock based experiments compare against 
the current bounds on fifth forces. The blue region shows the parameter 
space excluded by the MICROSCOPE experiment \cite{Berge2017_MICROSCOPE, 
Touboul:2017grn}. The red line depicts the projected sensitivity of an 
Earth based two-clock experiment involving a nuclear clock, the dark 
brown line shows the projected reach of the space based FOCOS experiment 
equipped with a nuclear clock, while the dark magenta and dark cyan lines 
project the sensitivities of the SpaceQ experiment assuming that the 
satellite reaches $r=0.39$AU and $r=0.1$AU respectively and is equipped 
with nuclear clocks. We take $\Delta K_{q} = 10^5$ and $\Delta 
\tilde{f}/\tilde{f} = 10^{-18}$ for all clock pairs except for the FOCOS experiment where we assume $\beta = 10^{-9}$. }
    \label{fig: EP experiments 3}
\end{minipage}%
\hfill
\begin{minipage}{.48\textwidth}
    \centering
    \includegraphics[width=0.995\linewidth]{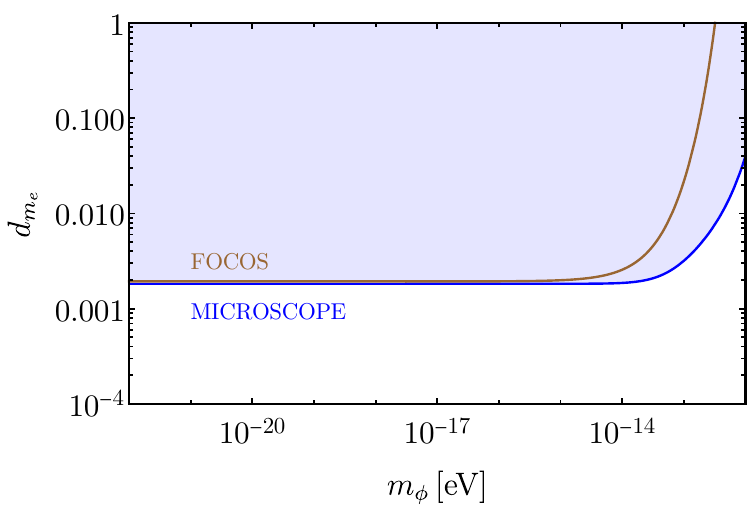}
    \caption{In the $d_{m_e}$ vs. $m_\phi$ plane we show how the 
projected limit from the future space based FOCOS experiment compares 
against the current bound. The blue region shows the parameter space 
excluded by the MICROSCOPE experiment 
\cite{Berge2017_MICROSCOPE,Touboul:2017grn} while the brown line shows 
the reach of the FOCOS experiment \cite{Derevianko:2021kye} assuming 
that the satellite is equipped with an optical clock with 
$K_{\alpha}=-6$ and the redshift is measured with the accuracy of $\beta = 10^{-9}$. \newline\newline\newline\newline\newline\newline  }
    \label{fig: focos dme}
\end{minipage}
\end{figure*}

\begin{figure}[h]
    \centering
    \includegraphics[width=0.995\linewidth]{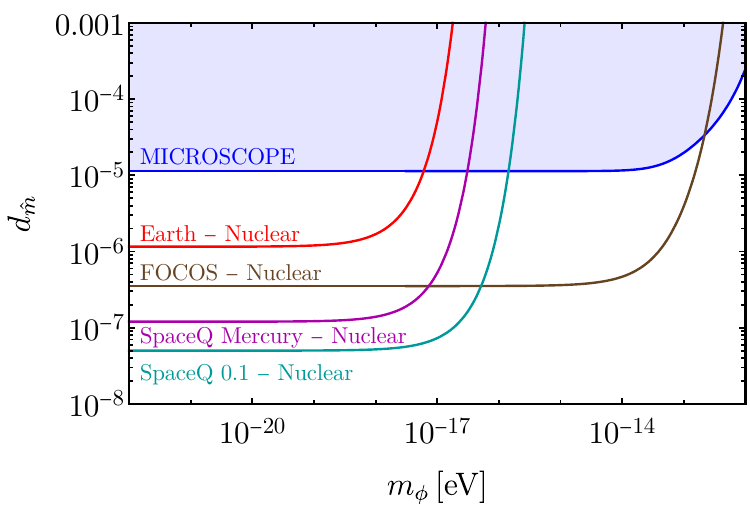}
    \caption{
   In the $d_{\hat{m}}$ vs. $m_\phi$ plane, we show how the projected 
limits from future clock based experiments compare against the current 
bounds from direct fifth force searches. The blue region shows the parameter 
space excluded by the MICROSCOPE experiment \cite{Berge2017_MICROSCOPE, 
Touboul:2017grn}. The red line gives the sensitivity of an Earth based 
two-clock experiment involving a nuclear clock, the dark brown line 
depicts the reach of the space based FOCOS experiment equipped with a 
nuclear clock while the dark magenta and dark cyan lines project the
sensitivities of the space based SpaceQ experiment assuming that the 
satellite reaches $r=0.39$AU and $r=0.1$AU respectively and is equipped 
with a nuclear clock. We assume $\Delta K_{q} = 10^5$ and $\Delta 
\tilde{f}/\tilde{f} = 10^{-18}$ for all clock pairs except for the FOCOS experiment where we assume $\beta = 10^{-9}$. }
    \label{fig: dmhat}
\end{figure}

\subsection{Two Different Clocks at the Same Location}

In order to match the sensitivity of the direct fifth force searches by 
the MICROSCOPE experiment~\cite{Berge2017_MICROSCOPE, Touboul:2017grn}, the uncertainty 
in the frequency of optical clocks located on earth needs to be of order 
${\Delta \tilde{f}}/{\tilde{f}} \sim 10^{-21}$ for $\Delta K_{\alpha} = 
7$. This requires an improvement by about 3 orders of magnitude over the 
best precision available at this time, which is expected to occur within the next two decades \cite{Derevianko:2021kye}.

However, improved precision is not the only option. Clocks based on 
nuclear transitions offer the exciting prospect of measuring the 
variation of fundamental constants with unprecedented sensitivity \cite{Peik:2020cwm}. The 
nuclear clock based on the $^{229}$Th nucleus is expected to have 
sensitivity to the variation of the fine structure constant about 3 
orders of magnitude better than the best optical clock, $K_{\alpha} \sim 
10^4$, while the sensitivity to the masses of the quarks is expected to 
be even greater, $K_{q} \sim 10^5$~\cite{Flambaum:2008kr}. Because of the large $K_q$, 
nuclear clocks of this type with an uncertainty of ${\Delta 
\tilde{f}}/{\tilde{f}} \sim 10^{-18}$ would be sufficient to improve on 
the current bounds set by the MICROSCOPE experiment. This is shown in 
Figs. \ref{fig: EP 
experiments 2}, \ref{fig: de_bounds}, \ref{fig: focos}
, and \ref{fig: EP experiments 3}.

Another possibility is to conduct an experiment in space. By sending the 
two clocks closer to the sun we can take advantage of the larger 
gravitational potential to increase the sensitivity to $(\beta_A - 
\beta_B)$. SpaceQ is a recent proposal based on this strategy 
\cite{Tsai:2021lly}. The satellite would be equipped with a two-clock 
system, and the frequency ratio between the clocks as the satellite 
orbits the sun would be measured. The first stage of the experiment 
proposes to send a satellite to Mercury's orbit, which is at $r=0.39$ 
AU. The second stage could reach down to $r=0.1$ AU. In both these 
stages the satellite's orbit is designed to be circular, which means 
that during the main stage of the experiment EP violating effects would 
not be measurable. However, as pointed out in the proposal, data can 
also be taken as the satellite transits towards its final orbit, 
allowing us to measure EP violating effects. In reaching the $r=0.1$ AU 
orbit, the satellite would experience a change in the gravitational 
potential of the order of $\Delta U \sim 10^{-7}$, which is greater than 
the annual modulation of the gravitational potential on earth by a 
factor of nearly 300. In Fig. \ref{fig: focos} we illustrate the 
sensitivity of this proposal to $d_e$ in four different scenarios. In 
the first case a satellite is sent out to Mercury's orbit containing two 
optical clocks. In the second case, one of the optical clocks is 
replaced by a nuclear clock. In the third and fourth cases a satellite 
is sent to orbit the sun at $r=0.1$ AU with an optical-optical and a 
nuclear-optical clock system respectively. In all versions of the 
experiment we assume a fractional uncertainty of ${\Delta 
\tilde{f}}/{\tilde{f}} \sim 10^{-18}$.  We see that the use of nuclear 
clocks leads to great improvements over the current sensitivity.

\subsection{Identical Spatially Separated Clocks}

The FOCOS experiment~\cite{Derevianko:2021kye} proposes to place a 
satellite carrying an optical clock in an elliptical orbit around the 
earth. The clock on the satellite will communicate with an identical 
clock on earth via optical links as the satellite approaches its apogee 
and perigee.  Since the distance between the surface of the earth and 
the satellite will vary between 5000 km and 22500 km, the satellite will 
experience a large variation in the earth's gravitational potential, 
allowing for an accurate determination of $\beta_A$. The optical clock 
offers more stability and precision than the microwave clocks that were 
used in earlier satellites. Instead of continuous monitoring of the 
frequency ratio, the experiment would monitor the phase difference 
between the clocks which can be translated into a frequency difference. 
This ultimately can be transformed into a limit on $\beta_A$ using 
Eq.~\eqref{eq: anomalous redshift experiment}. The experiment aims to 
measure $\beta_A$ with an accuracy of $10^{-9}$.

As explained in Sec.~\ref{sec: clock experiments}, the bound on the 
anomalous redshift can be translated into bounds on the parameters $D_x$ 
and $d_x$ through Eq.~\eqref{eq: beta A}. Since the proposal did not 
specify the clocks that would be used on the mission, we will consider 
two scenarios. In the first scenario, the satellite is equipped with an 
optical clock based on the electric octopole transition (E3) of 
$^{171}\text{Yb}^{+}$, which has the highest realized sensitivity to the 
variation of the fine structure constant, 
$K_{\alpha}=-6$~\cite{Flambaum:2008kr}. In the second scenario, we 
consider a satellite with a nuclear clock that has a sensitivity of 
$K_{\alpha} \sim 10^4$. The expected experimental reach of these two 
versions of the FOCOS experiment for the couplings $d_e$, $d_g$, 
$d_{m_e}$ and $d_{\hat{m}}$ is presented in Figs. \ref{fig: focos}, 
\ref{fig: EP experiments 3}, \ref{fig: focos dme} and \ref{fig: dmhat}. 
Remarkably, we see from Fig.~\ref{fig: focos dme} that with even with 
existing clock technology the FOCOS experiment would be competitive with 
the currents bounds on $d_{m_e}$ from direct fifth force searches. 
However, to improve on the current limits on $d_e$, $d_g$ and $d_{\hat{m}}$ would 
require the use of nuclear clocks.

It follows from this discussion that, when it comes to fifth forces 
mediated by scalars that couple primarily to electrons, satellite-based 
clock experiments can already compete with direct fifth force searches. 
The reason why this particular coupling is where clocks first start to 
gain ground is because the frequencies associated with most atomic 
transitions are directly proportional to $m_e$, and so they are very 
sensitive to changes in the electron mass.  In contrast, in direct fifth 
force searches, the electron only contributes a small amount to the mass 
of any given atom and so these experiments are inherently less sensitive 
to forces that act primarily on electrons.

\section{Conclusions}

In this article, we have explored how clock based experiments offer an 
alternative approach to probing EP violating fifth forces mediated by 
ultralight scalars. The same scalar field that mediates the fifth force 
will, in general, also give rise to position dependence in the 
fundamental parameters. Clocks on the earth as it orbits the sun or 
clocks on satellites orbiting the earth are sensitive to this effect, 
and therefore provide an excellent opportunity to test this class of 
models. The sensitivity of clock based experiments can be compared 
against the limits from direct fifth force searches, providing a 
benchmark to aim for when designing these experiments.

We have considered two classes of experiments utilizing clocks.  The 
first class of experiments we studied were differential measurements 
where two different clock transitions were being compared at the same 
location.  These experiments, performed on earth or in space, have the 
potential to probe beyond the current limits but require clocks more 
sensitive than currently available. The second class of experiments are 
anomalous redshift measurements where a clock in an elliptical orbit 
around the earth is compared to a clock on earth.  Experiments along 
these lines utilizing current clock technology can place constraints on 
fifth forces that act primarily on electrons that are competitive with 
current constraints.  Future nuclear clocks in elliptical orbits would 
offer a significant improvement in sensitivity to several different 
couplings.

The rapid experimental progress in clock technology has been quite 
remarkable.  Their sensitivity has been consistently improving by an 
order of magnitude every few years for the past several decades. Our 
analysis shows that if this rate of improvement is maintained, clock 
experiments may soon offer the greatest sensitivity to EP violating 
fifth forces mediated by ultralight scalar fields.

\section*{Acknowledgments}

We thank Marianna Safronova for useful discussions. ZC and 
IF are supported in part by the National Science Foundation under Grant 
Number PHY-1914731. ZC is also supported in part by the US-Israeli BSF 
Grant 2018236. AD is supported by the Fermi Research Alliance, LLC under Contract No. DE-AC02-07CH11359
with the U.S. Department of Energy, Office of Science,
Office of High Energy Physics. DB and AH are supported in part by the NSF under Grant 
No. PHY-1914480 and by the Maryland Center for Fundamental Physics 
(MCFP).

\appendix

\section{Effect of Uncertainty in the Earth's Gravitational Potential 
on Anomalous Redshift Measurements} 
 \label{appendix: deltaU}

 Fifth force searches based on anomalous redshift measurements require a 
knowledge of the general relativistic contribution to the redshift to 
the corresponding level of accuracy. Therefore uncertainties in our 
knowledge of the earth's gravitational field can limit the precision of 
the determination of the anomalous redshift. In this respect anomalous 
redshift measurements differ from direct fifth force searches or 
differential redshift measurements, for which the effect of the 
uncertainties in the earth's gravitational field is small because they
compare the motion of two bodies of different 
compositions or two different clocks at the same physical location.
In the neighborhood of the earth, the gravitational potential is 
determined from the precisely measured orbits of many different 
satellites under the assumption that they are acted on only by the 
gravitational force. However, as emphasized throughout this paper, 
models which give rise to position dependence of fundamental parameters 
also predict a fifth force.  This fifth force will affect the orbit of 
satellites and will therefore affect the inferred value of the 
gravitational potential. In this appendix, we systematically take this 
effect into account and show that the corrections to our formulas are 
small.

 Anomalous redshift experiments employ a clock placed on a satellite in 
orbit around the earth. The frequency of this clock is measured at 
different locations and compared against the frequency of an identical 
clock on earth. The anomalous redshift 
$\tilde{\beta}_A$ inferred from these experiments is related to the 
frequency change of the clock as,
 \bea \label{Eq: measured}
\frac{\Delta f}{f} = ( 1 + \tilde{\beta}_A) \Delta \tilde{U}_{\rm infer} \;,
 \eea
 Here $\tilde{U}_{\rm infer}$ is the inferred value of the difference in 
the gravitational potential as opposed to its actual value $\Delta 
\UGR$. In the class of theories we are exploring, this observed change 
in frequency actually arises from two separate contributions, one from 
general relativity and the other from the location dependence of 
fundamental constants.  Together, these read
 \bea 
\label{Eq: actual} 
\frac{\Delta f}{f} = \Delta \UGR + \beta_A \Delta \UGR \; , 
 \eea 
 where the first term is the difference predicted by general relativity 
while the second term is what was calculated in the text, see e.g. 
Eq.~(\ref{eq: clock comparison constants}).  To translate the experimental 
bound on $\tilde{\beta}_A$ to a constraint on $\beta_A$, we need to 
relate $\tilde{U}_{\rm infer}$ to $\UGR$.

The gravitational potential is inferred from the motion of satellites.  
The acceleration of a satellite in the earth's gravitational field is given
by,
 \bea \label{Eq: force}
a 
= \frac{G M}{r^2} \left ( 1 + \alpha_\text{\bf E} \alpha_\text{\bf S} \right ) 
= \frac{\Delta \UI}{r},
 \eea
 where $\alpha_{\bf E}$ and $\alpha_{\bf S}$ represent the values of 
$\alpha_{\bf X}$ for the earth and satellite respectively, where 
$\alpha_{\bf X}$ is as defined in Eq.~\eqref{eq: alpha_x}. For 
simplicity we have taken the mass of the ultralight scalar to be zero, since 
incorporating a non-zero mass for the modulus would require a detailed 
understanding of the orbits of the satellites employed.

 Using the relationship $\Delta \UGR = GM/r$ and Eqs.~\eqref{Eq: 
measured}, \eqref{Eq: actual} and~\eqref{Eq: force}, we find that
 \bea 
 \label{Eq: exact}
\tilde{\beta}_A \approx \beta_A - \alpha_{\bf E} \alpha_{\bf S} .
 \eea 
 From Eq.~\eqref{eq: beta A} we have
 \bea \label{eq: beta A for earth}
\frac{\beta_{A}}{\alpha_{\bf E}} = \left[(K^{A}_{\alpha}+2) d_{e} - K^{A}_{\mu} (d_{m_{e}}-d_g) + K^{A}_{q} (d_{\hat{m}}-d_g) + d_{m_{e}} \right] \; .
 \eea
 Then the inferred anomalous redshift $\tilde{\beta}_A$ is proportional
to the difference
  \bea 
\tilde{\beta}_{A} \propto F(K_X^A,d_x) - \alpha_{\bf S} \; ,
 \eea
 where $F(K_X^A,d_x)$ is the term on the right hand side of 
Eq.~\eqref{eq: beta A for earth}. From the expression for $\alpha_{\bf 
S}$ in Eqs.~\eqref{eq: alpha_x} and \eqref{eq: alpha_x_defs}, we see 
that this difference is the sum of a term proportional to $d_e$, a term 
proportional to $(d_{m_{e}}-d_g)$ and a term proportional to 
$(d_{\hat{m}}-d_g)$. As we now explain, this is exactly as expected. 
Recall that experiments cannot distinguish an EP preserving dilaton from 
general relativity in the non-relativistic limit. Setting the couplings 
to their values in the dilaton limit, $d_g = d_{m_e} = d_{\hat m}$ and 
$d_e = 0$, we find that $\tilde{\beta}_A = 0$ as expected. This 
constitutes a powerful cross check of our results.

From Eq.~\eqref{eq: alpha_x_defs} we see that the parameters 
$Q_{\hat{m}}$, $Q_{e}$ and $Q_{m_e}$ are all much less than one. 
This allows us to approximate
 \bea 
 \label{eq: beta_tilde A for earth}
\frac{\tilde{\beta}_{A}}{\alpha_{\bf E}} \approx \left[(K^{A}_{\alpha}+2) d_{e} + (1-K^{A}_{\mu})(d_{m_{e}}-d_g) + K^{A}_{q} (d_{\hat{m}}-d_g) \right]  \; .
 \eea 
 Comparing the right hand sides of Eqs.~\eqref{eq: beta_tilde A for 
earth} and \eqref{eq: beta A for earth}, we see that the difference 
between the two is just $d_g$. While the actual anomalous redshift 
scales with $d_g$ as $\beta_A/\alpha_{\bf E} \sim  (K^{A}_{\mu} - 
K^{A}_{q} ) d_g$, the inferred anomalous redshift scales as 
$\tilde{\beta}_{A}/\alpha_{\bf E} \sim  (K^{A}_{\mu} - K^{A}_{q} - 1) 
d_g$. Thus the errors arising from neglecting the effect of the fifth 
force on the motion of the satellite are small provided that either 
$d_g$ is negligible or $|K_{\mu}^A + K^{A}_{q}| \gg 1$. For the clocks 
used in the experiments we have considered, the second condition is 
satisfied.

\bibliography{EP_with_clocks}{}

\bibliographystyle{JHEP}
\end{document}